\documentclass[aps,prl,twocolumn]{revtex4}

\usepackage[english]{babel}
\usepackage{latexsym,amssymb,amsmath}

\begin{document}

\title{On the gauge coupling unification}
\author{E.K. Loginov}
\email{ek.loginov@mail.ru}

\affiliation{Department of Physics, Ivanovo State University, Ermaka St. 39, Ivanovo, 153025,
Russia}

\begin{abstract}
We consider a quark-lepton symmetry model of unification of the strong and electromagnetic
interactions. The model has the gauge group $SU(4)\times U(1)_{Y}$ and the minimal Higgs
structure consisting of one complex quartet of scalar particles. The spontaneous breakdown of
the gauge group to $SU(3)_{c}\times U(1)_{em}$ due to nonzero vacuum expectation value of the
Higgs quartet provides the simplest realization of the Higgs mechanism which generates masses
for gauge bosons, and masses to quarks and leptons. Using the embedding of the gauge group to
$SU(5)$, we study the evolution of coupling constants and find a connection of the couplings
with the gauge couplings of the standard model.
\end{abstract}
\maketitle

Despite the apparent striking success of the standard model (SM), there are a number of
properties that this model does not explain, including the quantization of the electric
charges of elementary particles, the ratios of the values of the respective standard-model
gauge couplings $g_{s}$, $g$, and $g'$, and the interconnected manner in which quark and
lepton contributions to gauge anomalies cancel each other (separately for each generation).
For the SM to be consistent the Higgs boson mass should be relatively light $M_{H}\leq 1$ TeV.
To explain the smallness of the Higgs boson mass some delicate cancellation is required that
is rather nontrivial ``fine tuning'' or gauge hierarchy problem. Other nontrivial problem is
that the SM can't predict the fermion masses, which vary over at least five orders of
magnitude (fermion problem). These and other deficiencies of the SM motivated the effort to
construct theories with higher unification of gauge symmetries.
\par
One of the possible variants of such theory can be induced by the possible four color
quark-lepton sym\-metry of Pati-Salam type treating leptons as the fourth color~\cite{pati74}
(see also~\cite{moha75}). In the present paper we propose a quark-lepton symmetry model of
unification of the strong and electromagnetic interactions. The model to be discussed here is
based on the $SU(4)\times U(1)_{Y}$ group as the minimal group containing the four-color
symmetry of quarks and leptons. Noting that $SU(4)\times U(1)_{Y}$ as a flavor group has been
considered before in the literature (see e.g.~\cite{volo88}). However, in such papers as a
rule possible extensions of the Glashow-Weinberg-Salam theory is studied. Whereas we construct
an analogue of the electroweak theory whose symmetry group $SU(4)\times U(1)_{Y}$ breaks down
to $SU(3)_{c}\times U(1)_{em}$.
\par
We start by the following observation~\cite{li74}. The gauge group $SU(5)$ has two
inequivalent maximal breaking patterns:
\begin{align}
SU(5)&\to SU(3)_{c}\times SU(2)\times U(1)_{Y},\\
SU(5)&\to SU(4)\times U(1)_{Y}.
\end{align}
In the first case, we obtain the Georgi-Glashow $SU(5)$-model~\cite{geor74a}. We shall study
the second case of the breaking patterns. As in the Georgi-Glashow model, we place the first
generation of fermions in the $\mathbf {5^{*}+10}$ representation of $SU(5)$ in the following
way:
\begin{equation}\label{52-01}
\begin{pmatrix} d^{c1}\\d^{c2}\\d^{c3}\\e\\-\nu
\end{pmatrix}_{L},\quad
\frac{1}{\sqrt{2}}
\begin{pmatrix}
0&u^{c3}&-u^{c2}&u_1&d_1\\
-u^{c3}&0&u^{c1}&u_2&d_2\\
u^{c2}&-u^{c1}&0&u_3&d_3\\
-u_1&-u_2&-u_3&0&e^{c}\\
-d_1&-d_2&-d_3&-e^{c}&0
\end{pmatrix}_{L}
\end{equation}
Now we note that the simplest representations of $SU(5)$ have the following $SU(4)\times U(1)$
content:
\begin{align}
5^{*}&=1(-2)+4^{*}(1/2),\label{52-02}\\
10&=4(3/2)+6(-1),\label{52-03}\\
24&=1(0)+15(0)+4(-2)+4^{*}(2).\label{52-04}
\end{align}
These representations uniquely (to within the chirality of fermions) determine multiplets of
fermions and gauge fields. Using (\ref{52-02}) and (\ref{52-03}), we get the following
multiplets of spinor fields:
\begin{equation}
e_{L},\quad\psi_{L}=\begin{pmatrix} d^{c1}\\d^{c2}\\d^{c3}\\-\nu
\end{pmatrix}_{L},\quad
\psi_{R}=\begin{pmatrix} u^{c1}\\u^{c2}\\u^{c3}\\-e
\end{pmatrix}_{R},
\end{equation}
and
\begin{equation}\label{52-11}
\chi=\frac{1}{\sqrt{2}}
\begin{pmatrix}
0&u^{c3}&-u^{c2}&d_1\\
-u^{c3}&0&u^{c1}&d_2\\
u^{c2}&-u^{c1}&0&d_3\\
-d_1&-d_2&-d_3&0
\end{pmatrix}_{L}.
\end{equation}
Using the adjoint representation (\ref{52-04}) we find multiplets of the vector fields:
sixteen massless gauge bosons of $SU(4)\times U(1)$, six gauge bosons $X'^{\alpha}_{\mu}$ and
$X'_{\mu\alpha}$ with the charges $\pm4/3$, and two gauge bosons $Y'^{\pm}_{\mu}$. The latter
eight gauge bosons obtain the masses via the Higgs mechanism.
\par
Now we write out the Lagrangian of $SU(4)\times U(1)_{Y}$-model. As in the
Glashow-Weinberg-Salam model, it can be divided to five  parts:
\begin{equation}\label{52-09}
\mathcal{L}
=\mathcal{L}_{YM}+\mathcal{L}_{HYM}+\mathcal{L}_{SH}+\mathcal{L}_{f}+\mathcal{L}_{Yuk},
\end{equation}
where
\begin{equation}
\mathcal{L}_{YM}=-\frac14F^{a}_{\mu\nu}F^{a\mu\nu}-\frac14G_{\mu\nu}G^{\mu\nu}
\end{equation}
is the Yang-Mills Lagrangian without matter fields,
\begin{equation}
\mathcal{L}_{HYM}=(D_{\mu}\varPhi)^{+}(D^{\mu}\varPhi)
\end{equation}
is the Lagrangian describing the Higgs quartet inter\-action with $SU(4)\times U(1)_{Y}$ gauge
fields,
\begin{equation}\label{52-08}
\mathcal{L}_{SH}=\mu^2\varPhi^{+}\varPhi-\lambda(\varPhi^{+}\varPhi)^2
\end{equation}
is the Lagrangian describing Higgs doublet self-inter\-action,
\begin{equation}\label{52-13}
\mathcal{L}_{f}=\bar\psi^{i}i\gamma^{\mu}(D_{\mu}\psi)^{i}
+\bar\chi_{ij}i\gamma^{\mu}(D_{\mu}\chi)_{ij}+\bar e_{L}i\gamma^{\mu}D_{\mu}e_{L}
\end{equation}
is the Lagrangian describing the interaction of fermions with gauge fields, and
\begin{multline}\label{52-12}
\mathcal{L}_{Yuk}=-f_{e}\bar\psi^{i}_{R}\varPhi^{i}e_{L}
-f_{u}\bar\psi^{i}_{R}\varPhi^{ijk}\chi_{jk}\\
-f_{d}\bar\psi^{i}_{L}\varPhi^{ijk}\chi_{jk}+h.c.
\end{multline}
is the Lagrangian describing the Yukawa interaction of the fermions with Higgs quartet.
\par
Here the covariant derivatives is given by
\begin{equation}\label{52-10}
D_{\mu}\varPsi=\left(\partial_{\mu}-ig\frac{\varLambda_{a}}{2}A^{a}_{\mu}
-ig'\frac{Y}{2}B_{\mu}\right)\varPsi,
\end{equation}
where values of $\varLambda_{a}$ dependent on the choice of the multi\-plet $\varPsi$. In
particular, if $\varPsi=e_{L}$, then all $\varLambda_{a}=0$. If $\varPsi=\chi$, then
$\varLambda_{p}=0$ as $p=1,\dots7$ and
\begin{equation}\label{52-06}
\varLambda_{s}=\frac{1}{\sqrt6}\begin{pmatrix}\lambda_{s}&0\\0&0
\end{pmatrix}
\end{equation}
as $s=8,\dots,15$. Here $\lambda_{s}$ are the usual Gell-Mann matrices for $SU(3)$. Finally,
if $\varPsi=\varPhi$ or $\varPsi=\psi\equiv\psi_{L}+\psi_{R}$, then
\begin{align}
\varLambda_{\alpha}&=\frac{1}{\sqrt6}(e_{4\alpha}+e_{\alpha4}),\label{52-05}\quad
\varLambda_{\alpha+3}=\frac{i}{\sqrt6}(e_{4\alpha}-e_{\alpha4}),\\
\varLambda_7&=\frac{1}{6}(e_{11}+e_{22}+e_{33}-3e_{44}),
\end{align}
and $\varLambda_{a}$ are defined by (\ref{52-06}) as $a=8,\dots,15$. Here $e_{ij}$ are
matrices with the elements $(e_{ij})_{mn}=\delta_{im}\delta_{jn}$ and $\alpha=1,2,3$.
\par
It follows from (\ref{52-10}) and (\ref{52-06}) that the multiplet (\ref{52-11}) must be
$SU(3)$-invariant. On the other hand, the skew-symmetric representation of $SU(4)$ has the
$SU(3)$ content $\mathbf{6=3+3^{*}}$. Therefore in (\ref{52-13}) and (\ref{52-12}) we use the
representation
\begin{multline}\label{52-14}
\chi=\frac{1}{\sqrt{2}}[(u^{c1}e_{[23]}+u^{c2}e_{[31]}+u^{c3}e_{[12]})_{L}\\
+(d^{c1}e_{[23]}+d^{c2}e_{[31]}+d^{c3}e_{[12]})_{R}],
\end{multline}
where $e_{[ij]}=e_{ij}-e_{ji}$, in place of (\ref{52-11}). Here we take advantage of the
charge conjugation formula and $SU(4)$-invariance of the Levi-Civita tensor
$\varepsilon^{ijkl}$.
\par
The hypercharge $Y$ is equal to the doubled mean charge of the multiplet
\begin{equation}
Y=2\langle Q\rangle,
\end{equation}
and also
\begin{equation}
Y_{\psi}=2Q-\varLambda_7.
\end{equation}
It follows from (\ref{52-12}) that
\begin{equation}
Q(\varPhi_{\alpha})=1/3,\quad Q(\varPhi_4)=0.
\end{equation}
Finally, the complex multiplet
\begin{equation}
\varPhi^{ijk}=\frac{1}{\sqrt2}\,\varepsilon^{ijkl}\varPhi_{l}.
\end{equation}
\par
For $\mu^2,\lambda>0$, the vacuum expectation values of the components of $\varPhi$ are
\begin{equation}\label{52-25}
\langle\Phi_{\alpha}\rangle=0,\quad \langle\Phi_0\rangle\equiv v=\frac{\mu}{\sqrt{\lambda}}.
\end{equation}
so that we have the breaking pattern
\begin{equation}
SU(4)\times U(1)_{Y}\to SU(3)\times U(1),
\end{equation}
In order that to define the mass spectrum we define the fields
\begin{equation}\label{52-15}
W'^{\alpha}_{\mu}=\frac{1}{\sqrt2}\left(A_{\mu}^{\alpha}-iA_{\mu}^{\alpha+3}\right)
\end{equation}
and diagonalize the mass matrix by the orthogonal transformation
\begin{align}
Z'_{\mu}&=A_{\mu}^7\cos\theta-B_{\mu}\sin\theta,\\
A_{\mu}&=A_{\mu}^7\sin\theta+B_{\mu}\cos\theta\label{52-16}
\end{align}
with
\begin{equation}\label{52-21}
\tan\theta=\frac{g'}{g}.
\end{equation}
Then we get
\begin{equation}\label{52-24}
M^2_{W'}=\frac{g^2v^2}{24},\quad M^2_{Z'}=\frac{v^2(g^2+g'^2)}{16}.
\end{equation}
\par
Substituting the fields (\ref{52-15})--(\ref{52-16}) in the Lagrangian (\ref{52-09}), we get
\begin{equation}\label{52-17}
\mathcal{L}=\mathcal{L}_{QCD}+\mathcal{L}_{CC}+\mathcal{L}_{NC}+\dots,
\end{equation}
where
\begin{equation}
\mathcal{L}_{QCD}=-\frac14F_{\mu\nu}^{k}F^{k\mu\nu}+\bar \psi_{q}^{c\alpha}\left(i\gamma^{\mu}
\tilde D_{\mu}+m_{u}\right)\psi_{q}^{c\alpha}.
\end{equation}
Here the anti-quarks $u^{c\alpha}$ and $d^{c\alpha}$ are denoted by the symbols
$\psi^{c\alpha}_{q}$,
\begin{equation}
\tilde D_{\mu}=\partial_{\mu}-ig\frac{\lambda_{s}}{2\sqrt6}A^{s}_{\mu},\quad
m_{\psi}=\frac{v}{\sqrt2}f_{\psi}.
\end{equation}
Suppose $m_{\psi}$ is the mass of $\psi$ and $g$ is connected with the effective QCD coupling
$\alpha_{s}$ by the relation
\begin{equation}\label{52-19}
g^2=24\pi\alpha_{s}.
\end{equation}
Then $\mathcal{L}_{QCD}$ is exactly coincides with the Lagrangian describing the strong
interactions of colored anti-quarks and gluons in the SM. The interaction of charged currents
is described by the Lagrangian
\begin{equation}
\mathcal{L}_{CC}=-\frac{g}{2\sqrt{3}}J_{\mu}^{\alpha}W'^{\alpha}_{\mu}+h.c.,
\end{equation}
where
\begin{equation}\label{52-29}
J_{\mu}^{\alpha}=\bar d^{c\alpha}_{L}\gamma_{\mu}\nu_{L}+\bar
u^{c\alpha}_{R}\gamma_{\mu}e_{R}.
\end{equation}
Obviously, the vector leptoquark $W'^{\alpha}_{\mu}$ has the electric charge $+1/3$. For the
neutral currents in (\ref{52-17}), we have
\begin{equation}\label{52-18}
\mathcal{L}_{NC}=g\sin\theta J^{\mu}_{em}A_{\mu}+\frac{g}{\cos\theta}J_0^{\mu}Z'_{\mu}.
\end{equation}
where
\begin{equation}
J^{\mu}_{em}=Q^{(q)}\bar\psi_{q}^{c\alpha}\gamma^{\mu}\psi_{q}^{c\alpha}
+Q^{(l)}\bar\psi_{l}\gamma^{\mu}\psi_{l},
\end{equation}
and
\begin{multline}
J^{\mu}_{0}=N_{L}^{(q)}\bar\psi_{qL}^{c\alpha}\gamma^{\mu}\psi_{qL}^{c\alpha}
+N_{R}^{(q)}\bar\psi_{qR}^{c\alpha}\gamma^{\mu}\psi_{qR}^{c\alpha}\\
+N_{L}^{(l)}\bar\psi_{lL}\gamma^{\mu}\psi_{lL} +N_{R}^{(l)}\bar\psi_{lR}\gamma^{\mu}\psi_{lR}
\end{multline}
Here the anti-quarks and leptons are denoted by the symbols $\psi^{c\alpha}_{q}$ and
$\psi_{l}$ respectively, $Q^{(f)}$ is a charge of the fermion $\psi_{f}$ and $N^{(f)}$ is a
diagonal element of the matrix
\begin{equation}
N=\frac{\varLambda_7}{2}-Q\sin^2\theta.
\end{equation}
Suppose that $g$ is connected with the effective QED coupling $\alpha$ by the relation
\begin{equation}\label{52-20}
g^2\sin^2\theta=4\pi\alpha.
\end{equation}
Then the first term in the right hand side of (\ref{52-18}) is exactly coincides with the
Lagrangian describing the electromagnetic interaction in the Glashow-Weinberg-Salam model.
Comparing (\ref{52-19}) and (\ref{52-20}), we get
\begin{equation}
\sin^2\theta=\frac{\alpha}{6\alpha_{s}}.
\end{equation}
\par
Now we again imbed the $SU(4)\times U(1)_{Y}$ gauge group in the $SU(5)$ gauge group and
consider the evolution of the coupling constants $g$ and $g'$. The evolution of couplings in
gauge theories with the groups $SU(n)$ is described by the renormalization group equation
\begin{equation}
\frac{dg_{n}}{d(\ln\mu)}=-b_{n}g^3_{n}.
\end{equation}
Solution of this equation for $g_{n}$ is
\begin{equation}\label{52-22}
g_{n}^{-1}(\mu)=g_{n}^{-1}(\mu_0)-\frac{b_{n}}{8\pi^2}\ln\frac{\mu_0}{\mu}.
\end{equation}
Since the $SU(4)\times U(1)_{Y}$ gauge theory has four multiplets of chiral fermions and one
quartet of charged scalars, it follows that the one-loop beta function coefficients
\begin{align}
b_4&=44/3-4n/3-1/6,\\
b_1&=-20n/9-1/6,
\end{align}
where $n$ is the number of generations. On the other hand, the embedding of $SU(4)\times
U(1)_{Y}$ in the $SU(5)$ gauge group imposes the coupling constant relations
\begin{equation}
g^2=15g'^2
\end{equation}
as $\mu=\mu_0$. Using (\ref{52-21}) and (\ref{52-20}), we represent the relation (\ref{52-22})
for $n=1$ and 4 in the form
\begin{align}
\alpha^{-1}(\mu)\sin^2\theta&=\alpha^{-1}_0-\frac{1}{2\pi}b_4\ln\frac{\mu_0}{\mu}\\
\frac{1}{15}\alpha^{-1}(\mu)\cos^2\theta
&=\alpha^{-1}_0-\frac{1}{2\pi}b'_1\ln\frac{\mu_0}{\mu},
\end{align}
where $b'_{1}=b_1/15$. Given the experimental inputs~\cite{amsl08}:
\begin{align}
\alpha^{-1}(M_{Z})&=127.909\pm 0.019,\\
\alpha_{s}(M_{Z})&=0.1217\pm 0.017,\\
M_{Z}&=91.1874\pm 0.0021\quad\mathrm{GeV}
\end{align}
we have
\begin{align}
\sin^2\theta&=0.01071\pm 0.00015,\\
\mu_0&=5248\pm 65\quad\mathrm{GeV}\label{52-23}.
\end{align}
\par
We assume that (\ref{52-23}) defines the symmetry-breaking scale of the $SU(4)\times U(1)_{Y}$
gauge theory, i.e. that $\mu_0$ is coincided with the vacuum expectation value $v$ in
(\ref{52-25}). Then we get
\begin{align}
M_{W'}&=3245\pm 63\quad\mathrm{GeV},\label{52-26}\\
M_{Z'}&=3996\pm 77\quad\mathrm{GeV}.
\end{align}
Further, we consider the vector current effective Lagran\-gian
\begin{equation}
\mathcal{L}_{\mathit{eff}}=-\frac{g}{12M^2_{W'}}J_{\mu}^{\alpha}\bar J^{\alpha\mu},
\end{equation}
in the limit of energies low compared to the $W'$ mass. Suppose that
\begin{equation}\label{52-28}
\frac{g^2}{12M^2_{W'}}=|\varepsilon|\frac{G_{F}}{\sqrt2},
\end{equation}
where $G_{F}$ is the conventional Fermi coupling constant, known from the muon decay rate to
have the value
\begin{equation}\label{52-27}
G_{F}=1.166367(5)\cdot 10^{-5}\quad\mathrm{GeV^{-2}}.
\end{equation}
Substituting (\ref{52-26}) and (\ref{52-27}) in (\ref{52-28}), we find the factor
\begin{equation}
|\varepsilon|=(2.205\pm 0.116)\times 10^{-3}.
\end{equation}
It is interesting that this value is coincided with the measured CP-violating quantity
$|\varepsilon_{K}|$. Note also that the CP-noninvariant interaction with coupling constants
$G'\sim 10^{-3}G_{F}$ has been investigated in~\cite{wolf64}.
\par
Now we consider the $SU(4)\times U(1)$ model with several generations. We rewrite the current
(\ref{52-29}) in the form
\begin{equation}\label{52-30}
J_{\mu}^{\alpha}=\bar\psi^{c\alpha}_{qL}\gamma_{\mu}V'^{ql}\psi_{lL}+\bar
\psi^{c\alpha}_{qR}\gamma_{\mu}V''^{ql}\psi_{lR},
\end{equation}
where $V$ and $V'$ are unitary mixing matrices. Further, suppose the generations of the
$SU(4)\times U(1)$ model and CM are coincided. The current that couples to the $W$ boson field
in the CM is
\begin{equation}\label{52-31}
J_{\mu}=\bar\psi^{\alpha}_{qL}\gamma^{\mu}V^{qq'}\psi^{\alpha}_{q'L}
+\bar\psi_{lL}\gamma^{\mu}\psi_{lL}.
\end{equation}
Comparing (\ref{52-30}) and (\ref{52-31}), we get either $V'=V^{+}$ and $V''=1$ or $V''=V^{*}$
and $V'=1$. We consider the case of two generations. In this case, $V$ is a $2\times2$ unitary
matrix. Such a matrix has 4 parameters: one is a rotation angle, and the other three are
phases. We can remove two phases by performing the change of variables on the quark fields.
Since the currents (\ref{52-30}) and (\ref{52-31}) contain both the particles and
antiparticles, we cannot remove the third phase. The same set of arguments can be made for the
theory with three generations. The final form of the matrix $V$ contains 3 angles and 3 phase.
In particular, it follows from here that the Cabibbo-Kobayashi-Maskawa matrix~\cite{cabi63} is
obtained from $V$ as a special case. Hence, the $SU(4)\times U(1)$ model can provide an
alternative description of the observed CP violation.

\end{document}